\documentclass[12pt]{iopart}                     
\usepackage{graphicx}
\usepackage{lmodern}
\usepackage{iopams}
\usepackage{hyperref}
\usepackage{color}

\begin{document}

\title{Oscillations in SIRS model with distributed delays}

\author{Sebasti\'{a}n Gon\c{c}alves$^1$, G. Abramson$^2$ and Marcelo F. C. Gomes$^1$}

\address{$^1$Instituto de F\'{\i}sica, Universidade Federal do 
Rio Grande do Sul,Caixa Postal 15051, 90501-970 Porto Alegre RS, Brazil}
\address{$^2$Centro At\'{o}mico Bariloche, CONICET and Instituto
  Balseiro, 8400 S. C. de Bariloche, Argentina}
\ead{sgonc@if.ufrgs.br}
\ead{abramson@cab.cnea.gov.ar}
\ead{marfcg@if.ufrgs.br}
\maketitle

\begin{abstract}
The ubiquity of oscillations in epidemics presents a long standing
challenge for the formulation of epidemic models. Whether they are
external and seasonally driven, or arise from the intrinsic dynamics
is an open problem. It is known that fixed time delays destabilize the
steady state solution of the standard SIRS model, giving rise to
stable oscillations for certain parameters values. In this
contribution, starting from the classical SIRS model, we make a
general treatment of the recovery and loss of immunity terms.  We
present oscillation diagrams (amplitude and period) in terms of the
parameters of the model, showing how oscillations can be destabilized
by the shape of the distributions of the two characteristic
(infectious and immune) times. The formulation is made in terms of
delay equation which are both numerical integrated and
linearized. Results from simulation are included showing where they
support the linear analysis and explaining why not where they do
not. Considerations and comparison with real diseases are presented
along.
\end{abstract}

\section{Introduction}
Many diseases that have affected and still affect humans come and go
with time in a well established way. Examples are plenty and fill the
bulletins of world and national health organizations. Measles, typhus
and cholera epidemic waves, just to cite a few, are even part of
mathematical biology books~\cite{anderson,murray}.  The common
denominator of such diseases is the cyclic natural history of them, in
which a susceptible subject can go to infected, then to removed, and
finally back to the susceptible state. However, the mere cyclic nature
of the disease does not grant an oscillatory behavior of its epidemic,
as can be exemplified by gonorrhea~\cite{grassly2005,grenfell2005}.
In 2009 a new variant of influenza A H1N1, dubbed \emph{swine flu},
appeared in the scene taking the media to discuss on the wave behavior
of influenza.  In which way do these oscillations arise in a
population, apparently synchronizing the infective state of many
individuals? Are they related to external driving causes, such as the
seasons? Or do they arise dynamically from the very natural history of
the disease? It is known that several causes can produce oscillations
in model epidemic systems: seasonal driving~\cite{abramson2002},
stochastic dynamics~\cite{risau2007,aparicio2001}, a complex network
of contacts~\cite{kuperman2001}, etc.

In every infectious disease several characteristic times clearly
appear in the dynamics. In a SIRS type disease (see
figure~\ref{timeline}) one has an infectious time (during which the
agent stays in category $I$, infected and infectious), and an immune
time (during which the agent stays in category $R$, recovered from
infection, and immune to re-infection until it returns to the
susceptible class $S$). A standard SIRS model uses the inverse of
these characteristic times as rates in mass-action equations, showing
damped oscillations toward the endemic state in most typical
situations.  However, the standard SIRS model does not exhibit
sustained oscillations for any value of the parameters. Would it be
possible for a deterministic SIRS model to sustain oscillations, and
if so how do the characteristic times relate to the period of the
epidemic?  These are some of the questions we address in this
contribution.

As observed by Anderson and May \cite{anderson} the mathematically
convenient treatment of the duration of the infection as a constant
rate is rarely realistic. It is more common that recovery from
infection takes place after some rather well defined time.
It would seem a valid simplification to assume that recovery happens
\emph{exactly} at a given time, instead of continuously at some rate.
Both extremes are called Type B and Type A recovery
respectively by Anderson and May. Most infections belong to an
intermediate type between these two extremes (but closer to Type B).
These intermediate types have an infectious time distributed with a shape
between that of an exponential (Type A) and that of a delta
distribution (Type B). An accurate model should implement the
distributions based on empirical data.  And as suggested by
Hoppenstaedt and others \cite{hopp75}, that could be done replacing the
simple constant rate term by an integral, leading to
integro-differential equations. Similar considerations can be made
about the transition from recovered to susceptible term.
It is almost thirty years ago that Hethcote \etal
\cite{hethcote1981} showed precisely that the SIRS model with fixed delay
in the recovered to susceptible transition presents stable
periodic solution for certain parameter values.  On the other side
they demonstrated that fixed delay in the recovery from infection term
does not have the same effect when the other one is considered in the
usual way. This implies that the immunity time plays the most
important role in the emerging oscillations as we will show in detail.
Recently, continuing with the work of Hethcote \etal, Taylor and
Carr \cite{taylor09} studied in detail the dynamics of the SIRS model
with temporary immunity, but considering that a fraction of the
population acquire permanent immunity. In other words, that is
equivalent to a distribution of immunity times made of two delta peaks
(at a finite time and at infinite). That makes the analysis of the
oscillations much more involved because of the extra parameter: the
fraction of individuals who became permanently removed.

Our contribution can be regarded as an extension of the work of
Hethcote \etal with time delays, in which we use arbitrary
distributions of both infectious and immune times. While doing this we
want to keep the problem as simple as possible in terms of parameters,
thus we avoid the use of vital dynamics.  We start by considering the
most extreme case: fixed delay in both terms (delta distributions).
This is the simplest mathematical case in terms of delays, and it is
close related with the first case studied by Hethcote et
al.\cite{hethcote1981}.  Then, we consider a mono-parametric family of
models in which the times are described by continuous distributions ---being
possible to go continuously from a Type B to a Type A model for example. In
other words we go from the SIRS model with deterministic delays to the classical
constant rates SIRS, taking all the intermediate situations in between.  In this
way, while we cover previous results, we can go further considering the most
general situation for a SIRS model.

For delayed models, we analyze the onset of sustained oscillations and
characterize them with the parameters of the system. Linear analysis,
together with numerical solutions of the nonlinear model, provide a
clear characterization of the phenomenon. Stochastic numerical
simulations provide further support to our analysis. Moreover, we show
the effect of the shape of the distributed delays on the stabilization
of the oscillations. Besides, the oscillations period, which
may play an important role in the design of intervention policies, are
shown to satisfy general rules in terms of the SIRS parameters.

\section{SIRS model with arbitrary recovery and loss of immunity dynamics}
As mentioned above, the usual formulation of a SIRS model implies
that recovery from infection, $I\to R$, proceeds at a rate which is
independent of the moment of infection. Also, the loss of immunity
$R\to S$ is also just proportional to the current sub-population and proceeds at
its own rate. The mathematical formulation of such a model is usually presented
in terms of differential equations as follows:
\numparts
\begin{eqnarray}
 \label{sirs}
   \frac{\rmd s(t)}{\rmd t}=-\beta\,s(t)\,i(t) + \frac{r(t)}{\tau_r},\\
   \frac{\rmd i(t)}{\rmd t}= \beta\,s(t)\,i(t) - \frac{i(t)}{\tau_i}, \\
   \frac{\rmd r(t)}{\rmd t}= \frac{i(t)}{\tau_i} - \frac{r(t)}{\tau_r},
\end{eqnarray}
\endnumparts
where $s(t)$, $i(t)$ and $r(t)$ stand for the corresponding fractions
of susceptible, infectious and recovered individuals in the population
($s(t)+i(t)+r(t)=1$). The parameters of the model are $\beta$, the
contagion rate per individual, and $\tau_i$ and $\tau_r$, the
characteristic infectious and immunity periods
respectively\footnote{Note on nomenclature: the {\em infectious} time
is frequently called as {\em recovery} time as well, because it marks
the passage from $I$ to $R$ which is the recovery from
infection. While we prefer the first name, in order to avoid
ambiguities with labels we associate them to the classes, i.e.:
$\tau_i$ is the time in the infective class, $\tau_r$ is the time in
the recovered class.}

The analysis of more general systems ---in which the recovery from
infection and loss of immunity processes obey more general and more realistic
dynamics--- is more involved, leading to non-local
integro-differential equations. Before proceeding to the most
general situation, we analyze the simplest case of \emph{fixed
times}.

\subsection{SIRS with fixed infectious and immunity times}
\begin{figure*}[htp]
\centering \includegraphics[width=8.4cm, clip=true]{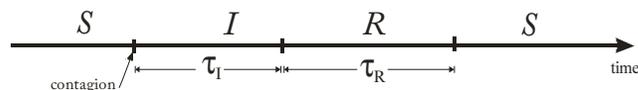}
    \caption{Timeline of an individual, showing the course of the disease
    after contagion.}
    \label{timeline}
\end{figure*}

Let us assume that the disease is characterized by an
\emph{infectious time} $\tau_i$ as well as an \emph{immunity time}
$\tau_r$. That is, an individual that becomes infectious at time $t$
will \emph{deterministically} recover at time $t+\tau_i$, becoming
immune, and will subsequently loose its immunity at time
$t+\tau_i+\tau_r \equiv t+\tau_0$, becoming susceptible again. The
process is schematically depicted in figure~\ref{timeline}.
This system can be represented by the following set of equations for
the fraction of susceptible and infectious sub-populations (bear in
mind that $r(t)=1-s(t)-i(t)$, so that just two equations describe
the dynamics):
\numparts
\begin{eqnarray}
\label{spunto} \frac{\rmd s(t)}{\rmd t} = -\beta\,s(t)\,i(t) + \beta\,s(t-\tau_0
)\,i(t-\tau_0 ), \\
\label{ipunto} \frac{\rmd i(t)}{\rmd t} =  \beta\,s(t)\,i(t) - \beta\,s(t-\tau_i
)\,i(t-\tau_i ).
\end{eqnarray}
\endnumparts

\begin{figure*}[htp]
\centering \includegraphics[width=8cm, clip=true]{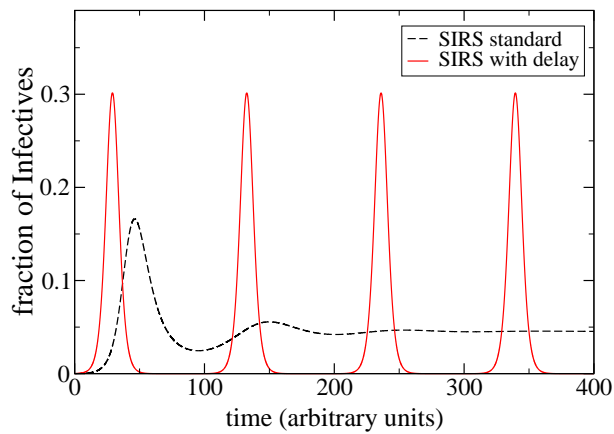}
    \caption{Time evolution of the fraction of infected individuals for
      the SIRS model. Comparison between the numerical solutions for
      the standard and the two fixed time delays formulation.
      Parameters are $\beta=0.4, \tau_i=5,  \tau_r=50$.}
    \label{SIRS}
\end{figure*}

Before proceeding with the detailed analysis of the above formulation
of the SIRS model we show in figure~\ref{SIRS} a result in advance,
comparing the time evolution of the fraction of infectives in the two
scenarios: the SIRS with two fixed time delays and the standard
SIRS. The last one produce the well know behavior of damped
oscillations toward the endemic state. Using the same parameters
($\beta=0.4, \tau_i=5, \tau_r=50$) the SIRS with two delays shows
clearly the sustained peaked oscillation in the infective fraction.
The time delay in the removed to susceptible transition instead of the
usual continuous rate transition is the responsible for the oscillation as we will see in the next sections. In general, delay equations
applied to SIR like models have noticeable effects on the dynamics,
that is different forms of the infection time distribution yield
different dynamics for the infectives and susceptibles~\cite{gomes2009}.
The example of figure~\ref{SIRS} is a remarkable case.

In equations~(\ref{spunto}-\ref{ipunto}), the first terms represents the contagion
of susceptible by infectious ones, which occurs locally in time at a
rate $\beta$. The second terms account for loss of infectivity
(\ref{ipunto}) and loss of immunity (\ref{spunto}). Both
terms correspond to the individuals infected at some earlier time:
$t-\tau_i$ and $t-\tau_0$ respectively, and who have proceeded through
the corresponding stage of the disease. These equations must
be supplemented with initial conditions, appropriate for interesting
epidemiological situations. A reasonable choice, which we use in the
remaining of the paper, is an introduction of infectious subjects into
a completely susceptible population:
\begin{equation}
\label{condin} s(0)=1-i_0,\,\, i(0)=i_0,\,\,r(0)=0.
\end{equation}

The system (\ref{spunto}-\ref{ipunto}) has the drawback that any pair of constants
$(s_0,i_0)$ satisfies them, apparently indicating that any pair of
values are equilibria. The origin of this problem lies in the fact
that (\ref{spunto}-\ref{ipunto}) together with (\ref{condin}) do not constitute a
well-posed differential problem.  Due to the non-locality in time,
extended initial conditions must be provided. Mathematically, it is
usual to provide arbitrary functions $s(t)$ and $i(t)$ in the interval
$[-\tau_0,0)$. From an epidemiological point of view, however, it is
more reasonable to provide just the initial conditions at $t=0$, and
complementary dynamics in the intervals $[0,\tau_i)$: no loss of
infectivity or immunity, just local contagion; and $[\tau_i,\tau_0)$:
transitions from $I$ to $R$ (the second term of (\ref{spunto}) being
absent), and the functions $s(t), i(t)$ already obtained by the
initial dynamics.

Indeed, this is the most reasonable choice for the numerical solution
of the system, and it is the one we have followed in the numerical
results shown below. For the analysis of equilibria, however, an
integral representation of the system results into a better-posed
problem and the difficulties for the calculation of the equilibria
disappear.

An integral equation equivalent to (\ref{ipunto}) is:
\begin{equation}
\label{iint} i(t) = c_1 + \beta\int_{t-\tau_i}^{t} s(u)\,i(u)\,\rmd u,
\end{equation} 
the interpretation of which is immediate: the integral sums over all
the individuals that got infected since time $t-\tau_i$ up to time
$t$. These are all the infectious at time $t$, since those infected
before have already recovered. The integration constant $c_1$ is, in
principle, arbitrary, but it is easy to see that it must be zero since
no other sources of infectious exist beyond those taken into account
by the integral term.

Complementing (\ref{iint}) it is convenient to write the equation
for $1-r=s+i$, which cancels out the first terms in (\ref{spunto}-\ref{ipunto}):
\begin{equation}
\label{rint} s(t)+i(t) = c_2 - \beta\int_{t-\tau_0}^{t-\tau_i}
s(u)\,i(u)\,\rmd u,
\end{equation} 
where, again, $c_2$ is an integration constant. In this case we have
$c_2=1$ since no other sources of $R$ exist.

The system (\ref{iint},\ref{rint}) can be solved for the equilibria of
the dynamics, $s^*$ and $i^*$. One obtains:
\begin{equation} s^* = \frac{1}{\beta\tau_i},\,\, 
i^* = \frac{\beta\tau_i -1}{\beta\tau_0}, \label{pfijo}
\end{equation} 
which coincides with the equilibria found numerically by integrating
(\ref{spunto}-\ref{ipunto}), and also corresponds to the same equilibria that can
be found in a constant-rate SIRS model (\ref{sirs}), where the
rates of recovery and loss of immunity are $1/\tau_i$ and $1/\tau_r$
respectively.

\subsection{SIRS with general distribution of infectious and immunity times}
The idea behind the fixed-time delays can be generalized to describe
more complex dynamics. Let us start with the infected individuals,
which is simpler. Consider a probability distribution function $G(t)$,
representing the probability (per unit time) of loosing infectivity at
time $t$ after having become infected at time 0. Observe that the
fixed-time dynamics is included in this description, when
$G(t)=\delta(t-\tau_i)$. $G(t)$ can be used as an integration kernel
in a delayed equation for the infectious. Indeed, the individuals that
got infected at any time $u<t$ and cease to be infectious at time $t$ are:
\begin{equation*}
\beta\int_0^t s(u)i(u)\,G(t-u) \rmd u,
\end{equation*}
so that the differential-delayed equation for $i(t)$ is:
\begin{equation}\label{ipunto_delay}
\frac{\rmd i(t)}{\rmd t} = \beta s(t)i(t) - \beta\int_0^t s(u)i(u)\,G(t-u) \rmd u,
\end{equation}
analogous to (\ref{ipunto}).

A second kernel $H(t)$ must be considered for the loss of immunity
process. The differential equation for susceptible can then be written
as:
\begin{equation}\label{spunto_delay}
\frac{\rmd s(t)}{\rmd t} = -\beta s(t)i(t)+\beta \int_0^t \left[\int_0^v s(u)i(u)\,G(v-u)\rmd u \right] H(t-v)
\rmd v,
\end{equation} 
where the second term corresponds to individuals that get infected at
earlier times, then loose their infectivity at intermediate times with
probability $G$, and finally recover at time $t$ with probability $H$.

The difficulty with initial conditions that we faced in the
fixed-times system is also found here, and can be solved in the same
way. Integral equations for $i(t)$ and $s(t)+i(t)$ result:
\begin{eqnarray}
\fl i(t) = c_1 + \beta\int_0^t s(u)i(u)\rmd u-\beta\int_0^t \left[ \int_0^v s(u)i(u)\,G(v-u)\rmd u \right] \rmd v, \\
\nonumber \fl s(t)+i(t) = c_2 + \beta\int_0^t\int_0^x \biggl[ \int_0^v s(u)i(u)G(v-u)\rmd u H(x-v) - \\
-s(v)i(v)G(x-v)\biggr] \rmd v\,\rmd x.
\end{eqnarray}

These equations can be used to find closed expressions for the
equilibria which, depending on the functional forms of $G$ and $H$,
can be solved analytically. In general one finds two sets of
solutions, the disease free one: $s^*=1$, $i^*=0$, and the endemic one
bifurcating from it:
\begin{equation} s^*=\frac{1}{\beta
\Sigma_1},~~i^*=\frac{\beta\Sigma_1 -1}{\beta (\Sigma_1-\Sigma_2)},
\label{pfijogen}
\end{equation} with:
\begin{equation} \Sigma_1 = \int_0^\infty \biggl[1-\int_0^v G(v-u)\rmd u
\biggr] \rmd v,
\end{equation}
\begin{equation} \Sigma_2 = \int_0^\infty \int_0^x
\biggl[H(x-v)\int_0^v G(v-u)\rmd u -G(x-v) \biggr] \rmd v\,\rmd x.
\end{equation}
We observe that using either Dirac deltas or exponential functions for
the kernels (corresponding to fixed-times and constant rates,
respectively), these integrals can be solved analytically to find the
equilibria.

Between the two extremes of constant rates and fixed delays, as
mentioned, realistic systems are expected to display some extended
probabilities distributions for infectious and immunity times. A
convenient interpolation between the exponential and delta
distributions that characterize those regimes can be achieved by gamma
distributions:
\numparts
\begin{eqnarray} 
 \label{gammas}
G_{p_i}(t) = \frac{p_i^{p_i} \,t^{p_i-1}
\rme^{-p_it/\tau_i}}{\tau_i^{p_i} (p_i-1)!}, \label{gp}\\
H_{p_r}(t) = \frac{p_r^{p_r} \,t^{p_r-1}\rme^{-p_rt/\tau_r}}{\tau_r^{p_r} (p_r-1)!}. \label{hp}
\end{eqnarray}
\endnumparts

These distributions have mean $\tau_i$ and $\tau_r$ respectively, for
any value of the parameter $p_{i,r}$. Besides, they interpolate
between exponential (when $p_{i,r}=1$) and Dirac delta distributions
(when $p_{i,r}\to\infty$), with smooth bell-shaped functions for
intermediate values of $p_{i,r}$. It can be shown that the equilibria
(\ref{pfijogen}) are identical to the classical ones (\ref{pfijo}) for
any $p_{i,r}$.

\section{Sustained oscillations in SIRS with delays}
Standard SIRS systems, without delays (or equivalently, with $G_1(t)$
and $H_1(t)$ as delay kernels) have either nodes or stable spirals as
equilibria. That is, oscillations appear in them as transient regimes
damped towards the fixed points. SIRS systems with delays, on the
other hand, can exhibit sustained oscillations. These appear as a Hopf
bifurcation of the spiral points, controlled by the parameters
$\tau_i$, $\tau_r$ and $\beta$. A linear stability analysis of the
fixed-times case can exemplify how this happens.

Assuming that the system (\ref{spunto}-\ref{ipunto}) is close to equilibrium, one
sets $s(t)=s^*+x(t)$, $i(t)=i^*+y(t)$, obtaining in linear
approximation a linear delay-differential system for the departures
from equilibrium:
\numparts
\begin{eqnarray}
\dot x(t)/\beta =-i^* x(t) -s^* y(t) +i^* x(t-\tau_0) + s^* y(t-\tau_0), \\
\dot y(t)/\beta =\;i^* x(t) +s^* y(t) -i^* x(t-\tau_i) - s^* y(t-\tau_i).
\end{eqnarray}
\endnumparts

\begin{figure*}[htp]
\centering \includegraphics[width=8.4cm, clip=true]{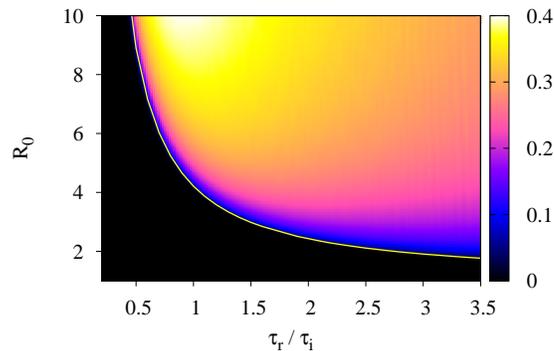}
    \caption{Bifurcation diagram of the SIRS model with fixed times in
the space defined by $\tau_r/\tau_i$ and $R_0$. The yellow(white) line
shows the linear result. The squared root amplitude of the infectives
oscillations is shown as colour (gray) coded shades above the
transition line. The black region means zero amplitude, representing
non-oscillatory endemic states.}
    \label{diag}
\end{figure*}

From this system, proposing exponential solutions $x(t)=c_1 \rme^{\lambda
t}$ and $y(t)=c_2 \rme^{\lambda t}$, a transcendental characteristic
equation is obtained:
\begin{equation}
\label{eqchar} \lambda^2
+\lambda\beta\left[s^*(\rme^{-\lambda\tau_i}-1)-i^*(\rme^{-\lambda\tau_{\,0}}-1)\right]
= 0.
\end{equation}
Equation~(\ref{eqchar}) can be solved numerically for complex
$\lambda$, obtaining from its real part the bifurcation line from the
stable spirals. This line is shown in figure~\ref{diag} along with the
amplitude of oscillations represented by a colour (gray) map.  The
amplitude is the result of the numerical integration of the full
nonlinear system~(\ref{spunto}-\ref{ipunto}).
Figure~\ref{diag} condenses the bifurcation phenomenon as a function
of the key parameters $\tau_r/\tau_i$ and $R_0$.  The black region
represents the non oscillating endemic solution.  It can be seen that
the linear analysis, represented by the line, defines almost exactly
the transition.

\subsection{Distributed delays: the general case}
The fact that we have qualitatively different results regarding the
nature of the endemic state for a delta or an exponential distribution
gives rise to a fundamental question. Is the existence of an
oscillatory endemic state particular to the delta distribution? Or is
there a critical shape of the distributions $G$ and $H$ necessary for
the emergence of the oscillations? Using the Gamma functions defined
in (\ref{gammas}) we can check the existence of such solutions for
different shapes by controlling the parameter $p_{i,r}$. In this way,
we can verify if there is a critical shape
$p_{i,r}=p_c(\beta,\tau_i,\tau_r)$ beyond which the system has
sustained oscillations at the endemic state. Linearizing the general
system (\ref{ipunto_delay},\ref{spunto_delay}) in the same way
presented in the previous section, one has the following integral
characteristic equation:
\begin{equation}\label{eqcarac} \fl \lambda^2+\lambda\beta i^*\left[1-
\int_0^{t}\int_0^{t-v}H(v)G(u)\rme^{-\lambda(u+v)}\rmd u\rmd v\right]-\lambda\beta
s^* \left[\int_{0}^{t}G(u)\rme^{-\lambda u} \rmd u\right]=0.
\end{equation}
Using the distributions (\ref{gammas}) in the equation above and
taking the limit $t\to\infty$ we get:
\begin{equation}\label{eqchar_p} \fl \lambda^2+\lambda\beta
i^*\left[1-\left(1+\frac{\lambda\tau_i}{p_i}\right)^{-p_i}\left(1+\frac{
\lambda\tau_r}{p_r}\right)^{-p_r}\right]-\lambda\beta
s^*\left[1-\left(1+\frac{\lambda\tau_i}{p_i}\right)^{-p_i}\right]=0.
\end{equation}
As expected, for $p_{i,r}=1$ and $p_{i,r}\to\infty$ we recover the
characteristic equations of the constant rates and fixed delays
models, respectively. Solving (\ref{eqchar_p}) numerically for
complex $\lambda$, one finds that for every set of parameters
(provided that $R_0=\beta\tau_i>1$) where $p_r>1$ there is always a
critical shape $p_i=p_c$ above which the endemic state consists of
sustained oscillations. Conversely, any value of $p_r>1$ can present
sustained oscillations in some region of parameter space.  Instead,
the constant rate model $p_r=1$ is a particular case where there is no
such solution for any set of parameters, as was demonstrated by
Hethcote \etal(\cite{hethcote1981}).

\begin{figure*}[thp] \centering \includegraphics[width=8cm,
clip=true]{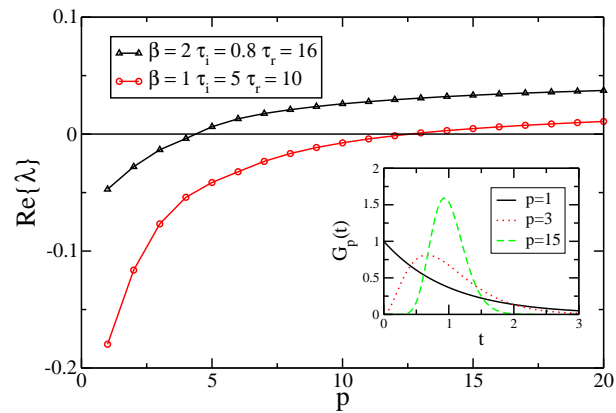}
\caption{Real part of the eigenvalue $\lambda$ as a function of the
shape parameter $p_r=p_i=p$. Inset: examples of the distribution $G_p$ with
$\tau_i=1$.} \label{relambda}
\end{figure*}

Figure~\ref{relambda} shows the real part of the solution of
(\ref{eqchar_p}) as a function of the shape ($p_{i,r}=p$) of the
two time distributions, and for fixed values of the parameters
$\beta,\,\tau_i$ and $\tau_r$. In other words we can appreciate (for
those $\beta,\,\tau_i$, $\tau_r$) the critical $p$ value ($p_c$) to
have oscillations in the SIRS dynamics, i.e. the value at which
$Re[\lambda]=0$.

\begin{figure*}[thp] \centering \includegraphics[width=8cm,
clip=true]{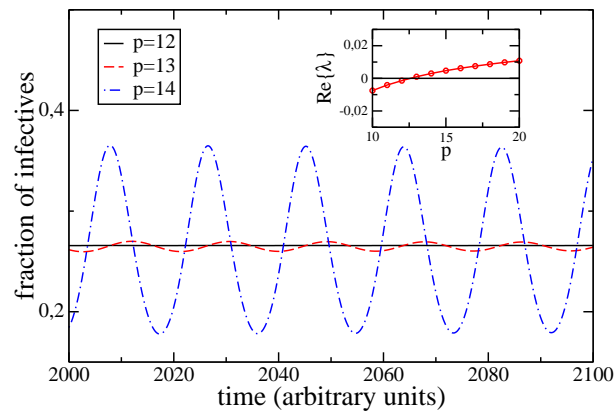}
\caption{Numerical integration of the SIRS general model with
$\beta=1,\,\tau_i=5,\,\tau_r=10$ near the critical shape $p_{i,r}=p_c$
predicted by the linear analysis of the system
(inset, $12<p_c<13$).} \label{oscilaciones}
\end{figure*}

The critical shape $p_c$ obtained by numerical calculation of
(\ref{eqchar_p}) gives an accurate prediction of the emergence of
oscillations in the full nonlinear system
(\ref{ipunto_delay},\ref{spunto_delay}), as can be seen in
figure~\ref{oscilaciones}. The meaning of the critical shape is simple:
the time distributions have to be narrower than the one represented by
$p_c$ in order to have sustained oscillations.

\begin{figure*}[thp] \centering \includegraphics[width=8cm,
clip=true]{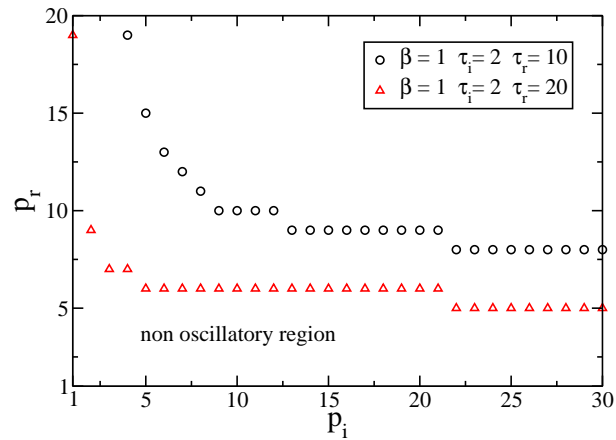}
\caption{Oscillation diagram in the $p_i$, $p_r$ plane for two sets of parameters:
$\beta=1$, $\tau_i=2$, $\tau_r=10$ (circles) and $\beta=1$, $\tau_i=2$, $\tau_r=20$ 
(triangles).} \label{pi-pr}
\end{figure*}

The distributions of infectious and immunity times used so far share a
common shape given by the value of $p$ ---that made the analysis,
restricted to only one shape parameter, simpler.  In real situations,
though, it is reasonable to expect that these uncorrelated kernels
have different shapes, not necessarily of the same relative width.
We explore this more general scenario, presenting an oscillation 
diagram in terms of $p_i$ and $p_r$ in figure~\ref{pi-pr}, for two sets
of the SIRS parameters.
Some interesting conclusion can be extracted from such diagram:
\begin{itemize}
\item
If the immunity time distribution is not sharp ($p_r<5$ for the
parameters used in figure~\ref{pi-pr}) there is no
oscillation independent of the type of infection time distribution.
\item
For a relatively narrow distribution of immunity times there is
a critical infective time width above which (i.e. a critical $p_i$
below which) no oscillations are obtained.
\item
The broader one of the time distribution is, the narrower the 
other one has to be in order to have oscillations.
\item
Longer immunity times (in units of infectious time) makes the oscillatory region wider in the $p_i, p_r$ space.
\end{itemize}

\begin{figure*}[thp] \centering \includegraphics[width=8cm,
clip=true]{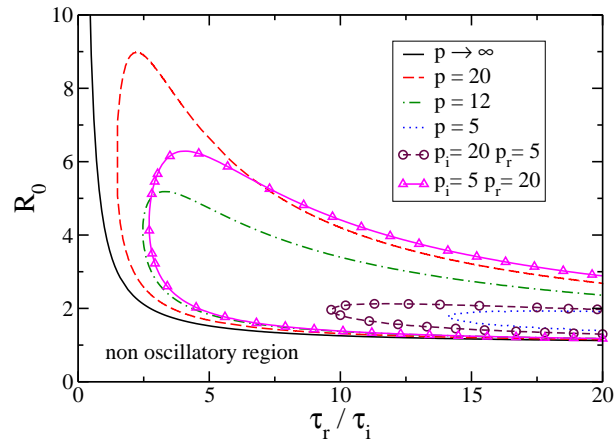}
\caption{Oscillation diagram in the $\tau_r/\tau_i$, $R_0$ plane for the SIRS
  model with different distribution for infectious and immunity times controlled
  by the $p_{i,r}$-shape factor of the Gamma distribution
  function. For each pair, $p_i$, $p_r$ the
oscillatory region is enclosed by the corresponding line. The curves
with only one $p$ are for $p_i=p_r=p$.}
\label{umbralp}
\end{figure*}

In the case $p_{i,r}\to\infty$ there is a critical value of
$R_0(\tau_r/\tau_i)$ above which the endemic solution is always a
stable cycle. The linear analysis also shows that for finite $p_{i,r}$
the bifurcation is more involved. Figure~\ref{umbralp} shows that the
bifurcation line \textit{encloses} a region of oscillating
solutions. Then, for a given value of $\tau_r/\tau_i$, there is a
second critical $R_0$, larger then the previous one, where the endemic
solution ceases to be cyclic. This phenomenon is verified in the
numerical solution of the nonlinear system.  For any given
$\tau_r/\tau_i$, if one increases $p$ the region of oscillation grows
(as shown in figure~\ref{umbralp}), so that in the limit
$p_{i,r}\to\infty$ the upper critical $R_0$ disappears. In such a
situation, the only way to break the oscillations is by decreasing
$R_0$. On the other hand, by decreasing $p_{i,r}$ the oscillatory
region shrinks, disappearing completely when $p_{i,r}=1$ (exponential
distributions, constant rates).

Yet, the interplay between the uncorrelated shapes $p_i$ and $p_r$ and
the SIRS parameters is very interesting. Again, in figure~\ref{umbralp}
we see for example that when $p_i=p_r=5$ (equal very broad shapes) the
oscillation region is very small. But increasing $p_i$ to 20 (narrow
infective times distribution), keeping the other fixed, expands the
region considerably. Similarly, but in other part of the same diagram,
when $p_i=p_r=20$ (equal very narrow shapes), the region of sustained
oscillations is relatively large. Now, by letting the distribution of
infective times to be broader ($p_i=5$) again with the other
distribution fixed, we see the oscillatory region to shrink
significatively.  So, while it is true that the immunity times
distribution must be necessarily different from an exponential in order
to have oscillations, in general both distributions are relevant to
define the actual parameter region of oscillations. The narrower one
or the two are, the wider the oscillatory region is.

\begin{figure*}[thp] \centering \includegraphics[width=8cm,
clip=true]{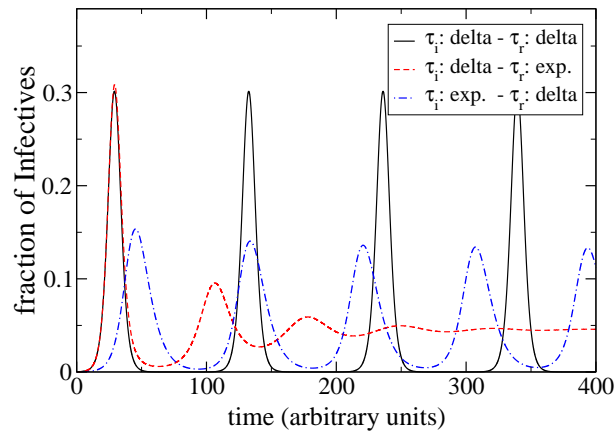}
\caption{Fraction of infectious as a function of time for three distributions of
  $\tau_r$ and $\tau_i$ with different shapes but same parameters $\beta=0.4$,
  $\tau_i=5$, $\tau_r=50$.}
\label{mix}
\end{figure*}

Three examples of dynamics of the SIRS with different shapes sets are
displayed in figure~\ref{mix}. The curves correspond to the numerical
solution of systems with the same epidemic parameters but three
different shape sets: ($p_i=p_r=\infty$), ($p_i=1, p_r=\infty$), and
($p_i=\infty, p_r=1$). The first two show the sustained oscillations
because the SIRS parameters are inside the oscillatory region, while
in the last case we know that there is no possible region of
oscillation so we have the damped behavior.  Besides, we see that
enlarging the $\tau_i$ distribution alone, while it shortens the
period of the oscillation, it does not destabilizes the
oscillation. Remarkably it shows an enhancement of the number of
infected during the low part of the cycles. This behavior diminishes
the probability of extinction in a discrete population realization of
the system, a problem that is common in simulations as we will see in
the next section.

\begin{figure*}[thp] \centering \includegraphics[width=8cm,
clip=true]{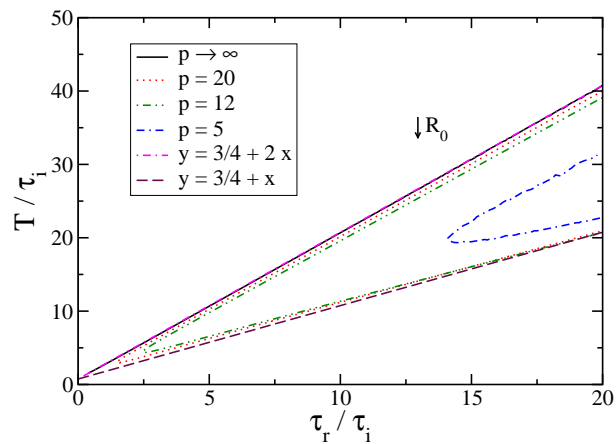}
\caption{Period of oscillations as a function of $\tau_r/\tau_i$. Each
curve corresponds to a different shape of the time distributions, as
shown in the legend where $p_i=p_r=p$. There are no oscillations above
and below the dashed lines.}
\label{periodo}
\end{figure*}

From the imaginary part of (\ref{eqchar_p}) it is possible to
obtain the period of oscillations in the linear approximation. In
figure~\ref{periodo} we show the result of this calculation for
different shapes of the distributions of infectious and immunity
times.  For finite $p_{i,r}$, the period of oscillation has a
dependency on the parameters that approaches the one found for
$p_{i,r}\to\infty$ (delta distributions, fixed times) in the lower
critical value of $R_0$, given by
\begin{equation} T=3/4 \tau_i + 2 \tau_r.
\end{equation} 
For a fixed value of $\tau_r/\tau_i$, this period decreases as $R_0$
increases up to the upper limit of oscillation, being bounded from
below by $3/4\tau_i+\tau_r$, as shown in figure~\ref{periodo}.  There
are no oscillations above and below these two straight lines, for any
value of $p_{i,r}$. This remarkable lock of the period of disease
oscillation is related to the one obtained by Taylor and
Carr (\cite{taylor09}) in the case of exponential distributed $\tau_i$
studied by them.

\section{Simulation}
\begin{figure*}[htp]
\centering \includegraphics[width=8.4cm, clip=true]{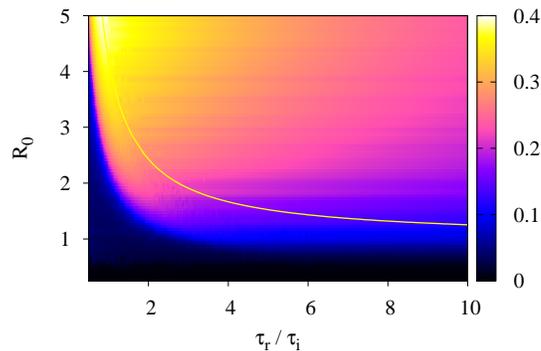}
    \caption{Diagram of the oscillation amplitude, in the
$\tau_r/\tau_i$, $R_0$ plane, obtained by simulation with $N=1000$
agents in the SIRS model with fixed times.  Vertical axis is $R_0 =
\beta \tau_i$, horizontal axis is the ratio $\tau_r/\tau_i$, and the
colour (gray) map represent the squared root amplitude of the $i(t)$ at
the steady state. The black region means zero amplitude, representing
non-oscillatory endemic or non-epidemic states. Superimposed is the
critical line obtained by the linear analysis as in figure~\ref{diag}.}
    \label{diagca}
\end{figure*}

\begin{figure*}[htp]
\centering \includegraphics[width=8cm, clip=true]{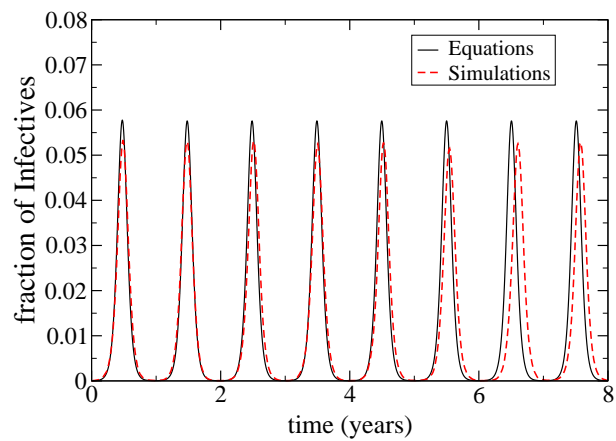}
    \caption{Evolution of an epidemic with stochastic and 
deterministic models. Time distributions
are deltas at $\tau_i=10 days$ and $\tau_r=180 days$, and $\beta = .13
per day$. With the time scale in years, the annual oscillation is 
clearly observed ($T = 2\tau_r + 3/4\tau_i=367 days$).}
    \label{eqxsim}
\end{figure*}

A complete picture of the general SIRS dynamics needs to contemplate
straightforward numerical simulations. We believe that simulations
represent the most accurate implementation of the real system, which
is discrete and stochastic. Therefore, as a test of our analytical
and numerical results with the generalized SIRS model, we present
here results from a probabilistic discrete model. In this
model a finite number of agents meet at random and contagion
proceeds in a probabilistic way. The phase diagram for such a
system, with delta distributions for $\tau_i$ and $\tau_r$, is shown
in figure~\ref{diagca}. The temporal evolution of the system is shown
in figure~\ref{eqxsim} along with the numerical solution of the
deterministic model.

\begin{figure*}[htp]
\centering \includegraphics[width=8cm, clip=true]{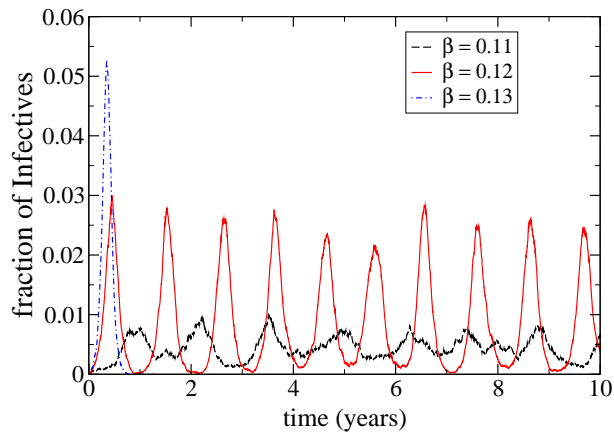}
    \caption{Evolution of the epidemic in the stochastic model, showing
extinctions of the infected population. Time distributions
are deltas at $\tau_i = 10$ and $\tau_r=180$ days, $N=10^5$.}
    \label{sim_beta}
\end{figure*}

The agreement between the two implementations of the same model is
very satisfactory, as can be seen by both the bifurcation diagram and
the temporal solution. Regarding the diagram the agreement is not as
perfect as with the numerical solution of the deterministic equations
(figure~\ref{diag}) because of fluctuations that arise in simulation due
to finite size. These cause large amplitude oscillations which become
extinct but that contribute with a non-zero value for the amplitude.
As for the temporal evolution, eventually, as time goes by, a small
drift develops in the simulation due to the stochastic nature of its
dynamics. Far from being disappointing this is a desirable feature,
since in real systems epidemic oscillations are not exactly periodic.
Yet, the extinctions, that are normally observed in the simulations,
are related with the discreteness of them.  Long time ago discussed by
Bartlett \cite{bartlett} and others it manifests here as being very
sensitive to the distance to the lower threshold in the bifurcation
diagram, which relates to the amplitude of the oscillation.  Greater
amplitudes drive the system closer to the absorbing state at $i=0$.
To illustrate this effect we show in figure~\ref{sim_beta} three
dynamics obtained by simulation with $N=10^5$ for three different
values of $\beta$, close to the onset of epidemics and oscillations.
However we saw in the previous section (figure~\ref{mix}) that a
distribution of infectious times with finite width can increase the
lower values of the infectives cycle, which eventually may prevent the
extinction in the simulation counterpart.

\section{Conclusions}
We have analyzed a general SIRS epidemic model, in which the infective
state as well as the immune state last prescribed times drawn from
distributions.  These distributed-time transitions constitute a
generalization of the standard SIRS model, in which the transitions
out of the infective and the immune states happen at a constant
rate. The generalization allows for situations in which these states
last for certain fixed times---which is more similar to many real
diseases than the constant rate assumption. Between the two extremes
of constant rates and fixed times, we have also analyzed the
intermediate situations of broader or narrower distributions of
the transition times.

The generalization runs along the proposals made by previous authors,
by implementing the differential equations of the model as non-local
in time. For example Hethcote etal. have shown that cyclic models have a
transition to oscillatory behavior when the immunity time is
distributed~\cite{hethcote1981}.

Our contribution shows how the oscillating state arises as a function
of all the parameters of the model, completing a phase diagram that
provides a thorough and general view of the possible behaviors. We
show, moreover, that the linear analysis, the numerical integration of
the model, and its stochastic implementation, all converge to the same
general picture, within the inherent limitations of each one.

For delta distributed delays (i.e., for fixed times spent in the
infective and recovered classes), the phase diagram of figure~\ref{diag}
shows that the region of oscillation is bounded from below by the
basic reproductive number $R_0$ as a function of the scaled immunity
time $\tau_r/\tau_i$. We see that, as $\tau_r$ decreases toward zero,
the minimum value of $R_0$ diverges. This result is in agreement with
the fact that in the SIS model there are no sustained oscillations. On
the other hand, the amplitude of the oscillations grows both with
$R_0$ and $\tau_r/\tau_i$. This kind of behavior may be relevant in
the analysis of real epidemics where changes in the parameters are
occurring due to interventions, advances in treatment or natural
causes.  For example, let us imagine an epidemic in the endemic
non-oscillating region (shaded black in the diagram), for which, as a
desirable consequence of the treatment of the disease, there is an
increase in the duration of the immune state, at constant $R_0$.  As a
result, the epidemic may start to oscillate. Such a transition may
manifest itself as an initial \emph{increase} of the infectious
fraction, due to the onset of oscillation, which should be properly
understood in the right context.

We have also shown that the transition to oscillations behaves as a
critical phenomenon depending on the \emph{width} of the time
distributions. Figure~\ref{relambda} shows this behavior. The
remarkable stabilization of the oscillations of pertussis after the
introduction of massive vaccination \cite[section 6.4.2]{anderson} 
may be related to a narrowing of the immunity time distributions.

For distributed delays, corresponding in our model to values of the
parameter $1<p_{i,r}<\infty$, we have found a remarkable reentrance
phenomenon in the phase diagram. As exemplified by figure~\ref{umbralp},
this reentrance means that the region of oscillations is also bounded
from above by a curve of $R_0$ vs $\tau_r/\tau_i$. One sees, also,
that the region of oscillation shrinks with lowering $p_{i,r}$, until
its disappearance at the critical value $p_c$. Only the case
$p_{i,r}\to\infty$, corresponding to delta distributed times is
unbounded. This is at variance with the SIR case analyzed by
\cite{black2009} who claims that there is no significant change
between large values of $p_{i,r}=p$ ($p=10$-$20$) and $p\to\infty$. On
the other hand our model supports their result in that the region of
oscillations is already large for $p$ in this range. An important
general result that can be derived from this diagram is the fact that,
whenever $\tau_r\gg\tau_i$, the minimum value of $R_0$ is very close
to 1, implying that such systems---provided that $p_{i,r}$ is
sufficiently large---are very prone to sustained oscillations. This
implies that oscillations due to the implementation of real time
distribution into the SIRS formalism are not a marginal or delicate
effect as someone could think before.

Our analysis also shows the behavior of the period of oscillation
and its dependence on system parameters. Is interesting to discuss
this aspect of our model in connection with real infectious
diseases. For example in the case of influenza, where the period of
oscillation is more or less a year and the infectious time is of the order of
10 days, our model predicts oscillations for any
$R_0>R_{0min}=1.09$. This minimum value is lower than the usual
estimate for this disease, which is around 1.4 \cite{hethcote2000}.
Indeed, this value is so close to $R_0=1$ that one can conclude that
there will always be oscillations for this disease as long as it
remains endemic. Besides we see that using a value of $\tau_r=180$
days for the loss of immunity, the region of oscillation lays
between $1.22<R_0<2.4$ for a rather wide distribution with $p_{i,r}=p=10$,
and $1.16<R_0<3.4$ for a narrower distribution with $p=30$. In the
first case, the corresponding period of oscillation is $188 < T < 342$
(in days), while on the second one it is $190 < T < 360$. Therefore, one
needs a narrow distribution to satisfy the observed period of
oscillation for influenza. In any case, the predicted region of oscillation
includes the observed values of $R_0$. We want to stress that, although
an infection by a specific strain of the influenza virus confers permanent immunity
after the infection period, the mutation rate of the virus can be thought of as an effective
immunity period and therefore can also be modeled with the SIRS dynamics.

In the case of other oscillating epidemics, such as syphilis and
pertussis, our model also fits the observed data rather well. For
syphilis, in the $p_{i,r}\to\infty$ case, using the period of 132
months measured in United States \cite{grassly2005} and the usual
value of $\tau_i=6$ months, we have an $R_{0min}=1.17$ and
$\tau_r\approx 6$ years. Greater values of $R_0$ (which are expected
for syphilis in ~\cite{grassly2005}), with fixed $\tau_i$, predict even
longer immune times (which is also the case of syphilis).  For
pertussis, using the period of 4 years and $\tau_i=1-2$ months, we
find $R_{0min}=1.06-1.15$, which is smaller than the estimations
obtained for mixing models~\cite{hethcote2000}.

In the open field of oscillatory epidemics, we believe that there is a
multiplicity of causes that can concur to produce the observed
phenomena. Our present contribution does not intend to provide an
exclusive and definitive answer to this matter.  The construction of
valid theoretical models for real diseases should incorporate all the
relevant mechanisms, therefore a thorough theoretical investigation of
any concurrent possible cause deserves due attention. We hope our
work has revealed one good candidate.

\ack
This work was supported by a cooperative agreement between Brazilian
Coordena\c{c}\~{a}o de Aperfei\c{c}oamento de Pessoal de N\'{i}vel Superior and Argentinian Ministerio de Ciencia y Tecnologia agencies (grant CAPES-MINCyT 151/08 - 017/07), and by Brazilian agency Conselho Nacional de Desenvolvimento Cient\'{i}fico e Tecnol\'{o}gico  (grant CNPq PROSUL-490440/2007). S.G. and M.F.C.G acknowledge financial support from CNPq, Brasil. 
G.A acknowledges support from Argentinian agencies Agencia Nacional de Promoci\'{o}n Cient\'{i}fica y Tecnol\'{o}gica (PICT
04/943), Consejo Nacional de Investigaciones Cient\'{i}ficas y T\'{e}cnicas (PIP 112-200801-00076), and Universidad Nacional de Cuyo (06/C304).

\end{document}